\documentclass[showpacs,pre,onecolumn]{revtex4}
\usepackage{graphicx}
\usepackage{subfigure}

\begin{document}

\title{Transitions between symmetric and asymmetric solitons in dual-core
systems with cubic-quintic nonlinearity}
\author{Lior Albuch and Boris A. Malomed}
\affiliation{Department of Interdisciplinary Studies, School of Electrical Engineering,
Faculty of Engineering, Tel Aviv University, Tel Aviv 69978, Israel}

\begin{abstract}
It is well known that a symmetric soliton in coupled nonlinear
Schr\"{o}dinger (NLS) equations with the cubic nonlinearity loses
its stability with the increase of its energy, featuring a
transition into an asymmetric soliton via a subcritical
bifurcation. A similar phenomenon was found in a dual-core system
with quadratic nonlinearity, and in linearly coupled fiber Bragg
gratings, with a difference that the symmetry-breaking bifurcation
is supercritical in those cases. We aim to study transitions
between symmetric and asymmetric solitons in dual-core systems
with saturable nonlinearity. We demonstrate that a basic model of
this type, \textit{viz}., a pair of linearly coupled NLS equations
with the cubic-quintic (CQ) nonlinearity, features a
\textit{bifurcation loop}: a symmetric soliton loses its stability
via a supercritical bifurcation, which is followed, at a larger
value of the energy, by a reverse bifurcation that restores the
stability of the symmetric soliton. If the linear-coupling
constant $\lambda $ is small enough, the second bifurcation is
subcritical, and there\ is a broad interval of energies in which
the system is \emph{bistable}, with coexisting stable symmetric
and asymmetric solitons. At larger $\lambda $, the reverse
bifurcation is supercritical, and at very large $\lambda $ the
bifurcation loop disappears, the symmetric soliton being always
stable. Collisions between solitons are studied too. Symmetric
solitons always collide elastically, while collisions between
asymmetric solitons turns them into breathers, that subsequently
undergo \textit{dynamical symmetrization}. In terms of optics, the
model may be realized in both the temporal and spatial domains.
\end{abstract}

\pacs{05.45.Yv; 42.65.Tg; 42.70.Nq}
\maketitle

\section{Introduction}

Twin-core nonlinear fibers and waveguides (alias \textit{couplers}) have
been a subject of considerable interest in nonlinear optics and, more
generally, in nonlinear-wave theory since the pioneering works by Jensen and
Maier \cite{Jensen}, who introduced the model of a dual-core nonlinear
optical fiber. In particular, many papers analyzed dynamics of solitary
waves in models of nonlinear couplers (see, e.g., review \cite{Wabnitz} and
Ref. \cite{Ivan}), although such solitons have not yet been created in the
experiment.

It has been known for a long time \cite{Wright} that a symmetric soliton in
the model of the twin-core fiber, with its energy equally divided between
the cores, becomes unstable when the energy exceeds a certain critical
value. Then, it was found that this instability gives rise to a \textit{%
pitchfork bifurcation}: a pair of new, stable \emph{asymmetric solitons}
(which are mirror images to each other) appear when the symmetric state
loses its stability \cite{CMP-Akhm}. This bifurcation was studied in detail
by means of numerical \cite{Akhm,Skinner} and analytical \cite{Skinner}
methods (the latter was based on the variational approximation, see Section
6 of review \cite{Progress}), with a conclusion that the bifurcation is
slightly \textit{subcritical}, according to the standard definition \cite%
{bif}. Namely, new stable asymmetric states appear at a value of the
soliton's energy which is slightly smaller than that at which the symmetric
soliton becomes unstable, thus giving rise to the \textit{bistability}
(coexistence of stable symmetric and asymmetric solitons) in a narrow
interval of energies before the pitchfork bifurcation.

The studies of solitons in nonlinear dual-core systems were extended in
various directions. First, the change of the character of the bifurcation
was investigated in a practically interesting case of a dual-core nonlinear
optical fiber with asymmetric cores \cite{Skinner,Dave}. Further,
four-component solitons were considered in a more sophisticated model, which
takes into regard two orthogonal polarizations of light in the dual-core
fiber \cite{Taras}. Then, bifurcations breaking the symmetry of
two-component solitons were studied in a symmetric dual-core nonlinear fiber
with the Bragg grating carried by each core \cite{Mak-Bragg}, and in a
system of parallel-coupled waveguides with the quadratic, $\chi ^{(2)}$
(rather than cubic, $\chi ^{(3)}$), nonlinearity \cite{Mak-Chi2}. In the two
latter cases, the symmetry-breaking bifurcation was found to be
supercritical, rather then subcritical (i.e., the destabilization of the
symmetric soliton immediately gives rise to a pair of stable asymmetric
ones, without the bistability). In addition, three-component solitons and
their bifurcations were studied in a system of three linearly-coupled fibers
forming a triangular configuration \cite{AkhmBur}, as well as in a similar
configuration formed by three Bragg gratings \cite{Arik}.

A universal feature of all the above-mentioned systems is that the symmetric
soliton is destabilized with the increase of its energy, as the
self-focusing nonlinearity favors states in which a larger part of the
energy is concentrated in one core (which is explained by the fact that the
part of the system's Hamiltonian accounting for the nonlinearity tends to be
lower in such an asymmetric configuration than in its symmetric
counterpart). Past the symmetry-breaking bifurcation point, the asymmetry of
the energy distribution between the cores increases monotonously with the
further increase of the total energy.

However, the self-focusing nonlinearity may feature saturation with the
increase of the power in some optical media. In particular, it was inferred
from experimental data that nonlinearity of chalcogenide glasses \cite%
{CQglass} and some organic transparent materials \cite{CQorganic} is
adequately approximated by adding a self-defocusing $\chi ^{(5)}$ (quintic)
term to the self-focusing $\chi ^{(3)}$ one. The respective \textit{%
cubic-quintic} (CQ) nonlinearity furnishes a paradigmatic example of the
saturation, as well as of the competition between self-focusing and
self-defocusing nonlinearities \cite{review}.

A model of a dual-core optical fiber or waveguide with saturable
nonlinearity is an interesting object in its own right, and may also be
relevant for applications to all-optical switching. In particular, the
symmetry-breaking bifurcation between continuous-wave (CW) states (rather
than solitons) in a model of a coupler with saturable nonlinearity was
considered long ago in Ref. \cite{Snyder}, with a conclusion that the
bifurcation is subcritical in this case. However, CW states in couplers with
a self-focusing nonlinearity are always subject to modulational instability
\cite{MI}, therefore it is necessary to consider solitons, rather than the
CW. Qualitatively new features may be expected in such systems. Indeed,
since the spontaneous symmetry breaking of soliton states, observed at
energy exceeding the critical value, is accounted for by the trend to
concentrate a larger part of the energy in one core with the growing
strength of the self-focusing, the saturation, which \emph{attenuates} the
self-focusing with the further increase of the energy, should eventually
give rise to a reverse trend. Thus, a \emph{reverse bifurcation} may occur
in the system, restabilizing the symmetric soliton and eliminating
asymmetric states.

This work aims is to produce a full picture of bifurcations in the symmetric
dual-core system with the CQ nonlinearity. In Section II, we introduce the
model, whose single free parameter is the constant $\lambda $ of the linear
coupling between the cores. Main results are presented in Section III: if $%
\lambda $ is smaller than a certain value $\lambda _{\max }$, the system
features a \textit{bifurcation loop}, first destabilizing and then
restabilizing the symmetric soliton, as may be expected (see above). The
direct (symmetry-breaking) bifurcation is always supercritical, while the
reverse one is subcritical (featuring a large bistability region, which may
be of considerable interest to applications \cite{Wright} - \cite{Ivan}),
except for a narrow interval near $\lambda =\lambda _{\max }$, where the
reverse bifurcation is supercritical, without the bistability. The loop
disappears at $\lambda =\lambda _{\max }$; in the region of $\lambda
>\lambda _{\max }$, the symmetric soliton is always stable, while asymmetric
ones do not exist. Direct simulations demonstrate that the stability of all
soliton branches exactly complies with what may be expected from general
principles of the bifurcation theory \cite{bif}.

In Section IV, we additionally investigate collisions between moving
solitons, which a natural issue to address as the Galilean invariance of the
model allows one to generate moving solitons. We infer from simulations that
collisions between stable symmetric solitons are completely elastic, while
asymmetric solitons collide inelastically and separate after the collision
in the form of breathers. Conclusions are formulated in Section V.

\section{The model}

Linearly coupled CQ nonlinear Schr\"{o}dinger (NLS) equations for the
amplitudes $u$ and $v$ of the electromagnetic fields propagating along the
coordinate $z$ in the two cores of an optical waveguide can be cast in a
normalized form,
\begin{eqnarray}
iu_{z}+u_{\tau \tau }+2|u|^{2}u-|u|^{4}u+\lambda v &=&0,  \label{u} \\
iv_{z}+v_{\tau \tau }+2|v|^{2}v-|v|^{4}v+\lambda u &=&0,  \label{v}
\end{eqnarray}
where $\tau \equiv t-z/V_{0}$ is the local time ($V_{0}$ is the group
velocity of the carrier wave). The model assumes anomalous group-velocity
dispersion, GVD (otherwise it will generate no bright solitons), whose
coefficient is normalized to be $1$. The coefficients in front of the
self-focusing $\chi ^{(3)}$ and self-defocusing $\chi ^{(5)}$ terms can
always be normalized as set in Eqs. (\ref{u}) and (\ref{v}), the only
remaining irreducible parameter of the model being the linear-coupling
constant $\lambda $ (we define it to be positive), that determines a
characteristic coupling length, $z_{\mathrm{coupl}}\sim 1/\lambda $ (in real
couplers, $z_{\mathrm{coupl}}$ takes values from $\sim 1$ mm up to $\sim 1$
cm, in physical units). In nontrivial soliton states (those which are
essentially affected by the linear coupling), the characteristic
nonlinearity and dispersion lengths, $z_{\mathrm{nonlin}}$ and $z_{\mathrm{\
disp}}$, must match $z_{\mathrm{coupl}}$. The successful creation of
solitons in fiber Bragg gratings, with the nonlinearity length $_{\sim
}^{<}~1$ cm \cite{BGsoliton}, suggests that $z_{\mathrm{nonlin}}$ for
high-power pulses in silica fibers can indeed be brought down to the
necessary size, in the range of $1$ mm -- $1$ cm (the nonlinearity length is
determined by the core itself, rather than by its coupling to another core
or grating written on its surface). On the other hand, while the intrinsic
GVD of the fiber waveguide may not be sufficient to provide for the
necessarily small $z_{\mathrm{disp}}$, the effective GVD may be enhanced by
several orders of magnitude by means of the same grating \cite{BGsoliton}.

In addition to the above-mentioned interpretation in the \textit{temporal
domain}, the model based on Eqs. (\ref{u}) and (\ref{v}) may also be
realized in terms of the light transmission in the \textit{spatial domain}.
In that case, the temporal variable $\tau $ is replaced by the transverse
coordinate $x$, and the equations govern the spatial evolution of amplitudes
of the electromagnetic field in a pair of parallel tunnel-coupled planar
waveguides, the terms $u_{xx}$ and $v_{xx}$ accounting for the transverse
diffraction of light (in the usual paraxial approximation). Launching the
beams with the transverse width $\Delta x\sim 20$ wavelengths (if the latter
is taken, as usual, in a ballpark of $\lambda \simeq 1$ $\mu $m; note that
the paraxial approximation is relevant for $\Delta x\sim 20\lambda $) will
provide for a value of the diffraction length, $z_{\mathrm{diffr}}=4\pi
\left(\Delta x\right) ^{2}/\lambda $, in the above-mentioned range of $1$ mm
-- $1$ cm, where it can be matched to the coupling length (for the parallel
planar waveguides, $z_{\mathrm{coupl}}$ is essentially the same as for the
dual-fiber system).

The starting point of analysis is a well-known exact soliton solution of the
single CQ NLS equation \cite{Pushkarov}, to which Eqs. (\ref{u}) and (\ref{v}%
) reduce in the symmetric case:
\begin{eqnarray}
u &=&v=e^{ikz}U_{\mathrm{symm}}(\tau ),  \nonumber \\
U_{\mathrm{symm}}(\tau ) &=&\sqrt{\frac{2\left(k-\lambda \right) }{1+\sqrt{
1-4\left(k-\lambda \right) /3}\cosh \left(2\sqrt{k-\lambda }\tau \right) }} ,
\label{Pushk}
\end{eqnarray}
where the propagation constant $k$ takes values in the interval $\lambda <k<%
\frac{3}{4}+\lambda $. In more sophisticated settings, solutions for a
single equation with the CQ nonlinearity can be found in a numerical form.
An example relevant to the spatial-domain model is a solution for the
semi-nonlinear equation, with an overall nonlinear coefficient (one in front
of both the cubic and quintic terms) present in the region of $x>0$ and
vanishing at $x<0$ \cite{Dum}.

Following the analysis of coupled NLS equations with the $\chi ^{(3)}$
nonlinearity \cite{Akhm,Skinner}, we look for asymmetric stationary
solutions to Eqs. (\ref{u}) and (\ref{v}) in the form of
\begin{equation}
\left\{ u(z,\tau ),v(z,\tau )\right\} =e^{ikz}\left\{ U(\tau ),V(\tau
)\right\} .  \label{uv}
\end{equation}
If the functions $U$ and $V$ are complex,
\begin{equation}
\left\{ U(\tau ),V(\tau )\right\} =\left\{ a(\tau )e^{i\phi (\tau )},b(\tau
)e^{i\psi (\tau )}\right\} ,  \label{complex}
\end{equation}
with real amplitudes and phases $\left\{ a,b\right\} $ and $\left\{ \phi
,\psi \right\} $, a straightforward corollary of the equations is the
following relation, valid for localized solutions (see details in Appendix):
\begin{equation}
a^{2}\dot{\phi}+b^{2}\dot{\psi}=0,  \label{ab}
\end{equation}
with the overdot standing for $d/d\tau $. For symmetric solitons, with $a=b$
and $\phi =\psi $, Eq. (\ref{ab}) implies, as usual, that all solutions are
real. While the same is not obvious for asymmetric solitons, we demonstrate
in Appendix that asymmetric complex solutions for $U(\tau )$ and $V(\tau )$
with a nontrivial phase structure cannot be generated by a bifurcation from
the symmetric soliton (\ref{Pushk}). For this reason, in what follows below
we consider real stationary solutions for $U$ and $V$, hence the
substitution of expressions (\ref{uv}) in Eqs. (\ref{u}) and (\ref{v}) leads
to a system
\begin{eqnarray}
\frac{d^{2}U}{d\tau ^{2}}-kU+\lambda V+2U^{3}-U^{5} &=&0,  \label{U} \\
\frac{d^{2}V}{d\tau ^{2}}-kV+\lambda U+2V^{3}-V^{5} &=&0.  \label{V}
\end{eqnarray}

We solved Eqs. (\ref{U}) and (\ref{V}) numerically by means of the
finite-difference method. In this paper, we only aim to search for
fundamental solitons, which correspond to even single-humped solutions to
these equations. Stability of the analytical symmetric solutions (\ref{Pushk}%
) and asymmetric ones found in the numerical form was tested by direct
simulations of Eqs. (\ref{u}) and (\ref{v}), performed by means of the
split-step method.

Note that Eqs. (\ref{u}) and (\ref{v}) conserve the total energy (norm) of
the solution,
\begin{equation}
N=\frac{1}{2}\int_{-\infty }^{+\infty }\left[ |u(\tau )|^{2}+|v(\tau )|^{2} %
\right] d\tau  \label{E}
\end{equation}
(in the spatial-domain optical model, the norm has the physical meaning of
the total power of the beams), together with the momentum and Hamiltonian
(that will not be used below). For exact solutions (\ref{Pushk}), the norm
is
\begin{equation}
N_{\mathrm{symm}}=\frac{\sqrt{3}}{2}\ln \left(\frac{\sqrt{3}+2\sqrt{
k-\lambda }}{\sqrt{3}-2\sqrt{k-\lambda }}\right) .  \label{Nsymm}
\end{equation}
It is worthy to note that expression (\ref{Nsymm}) satisfies the condition $%
dN/dk>0$, which, according to the known Vakhitov-Kolokolov (VK) criterion
\cite{VK}, is necessary for stability of the soliton family (the validity of
this criterion was established, in a general form, for single-component NLS
equations \cite{Luc}). The VK criterion guarantees the absence of unstable
modes of small perturbations with a real instability growth rate, but it
does not tell anything about modes corresponding to complex eigenvalues.

\section{The bifurcation diagram}

Symmetry-breaking bifurcation points of Eqs. (\ref{U}) and (\ref{V})
(critical values of $k$ for given $\lambda $) may be looked for by adding an
infinitesimal antisymmetric perturbation, $W(\tau )$, to the symmetric
soliton, $\left\{ U(\tau ),V(\tau )\right\} =$ $U_{\mathrm{symm}}(\tau )\pm
W(\tau )$. The substitution of this in Eqs. (\ref{U}) and (\ref{V}) and
linearization with respect to $W$ lead to a linear Schr\"{o}dinger equation
for $W$,
\begin{equation}
-\left(\lambda +k\right) W=-\frac{d^{2}W}{d\tau ^{2}}+U_{\mathrm{eff}}(\tau
)W,  \label{Linear}
\end{equation}
with an effective potential
\begin{equation}
U_{\mathrm{eff}}(\tau )\equiv U_{\mathrm{symm}}^{2}(\tau )\left[ 5U_{\mathrm{%
\ symm}}^{2}(\tau )-6\right] .  \label{Ueff}
\end{equation}
The first problem is to find values of $k$ at which an even localized state
appears among solutions of Eq. (\ref{Linear}). In the model with the cubic
nonlinearity only, Eq. (\ref{Ueff}) yields the effective potential $%
-6(k-\lambda )\mathrm{sech}^{2}(\sqrt{k-\lambda }\tau )$, which is
integrable, as is known from quantum mechanics. Therefore, the (single)
bifurcation point in the model of the coupler with the cubic nonlinearity
could be found exactly \cite{Wright}. In terms of the norm of the symmetric
soliton, it is $N_{\mathrm{cubic}}^{\mathrm{(bif)}}=2\sqrt{2\lambda /3}$.
However, the effective potential in the CQ model, given by Eqs. (\ref{Ueff})
and (\ref{Pushk}), is far from being integrable, therefore we will identify
bifurcation points by numerical methods.

Numerical solution of stationary equations (\ref{U}) and (\ref{V}) produces
a sequence of bifurcation diagrams displayed in Figs. \ref{bifurcation1} and %
\ref{bifurcation2}. In these diagrams, the soliton's \textit{asymmetry
parameter},
\begin{equation}
\epsilon \equiv \frac{U_{\max }^{2}-V_{\max }^{2}}{U_{\max }^{2}+V_{\max
}^{2}},  \label{epsilon}
\end{equation}
where $U_{\max }$ and $V_{\max }$ are the amplitudes of the two components
of the soliton, is shown versus the norm $N$. Note that the soliton's
propagation constant $k$ varies along with $N$ (not shown in the diagrams).
\begin{figure}[tbp]
\includegraphics[width=5.4in]{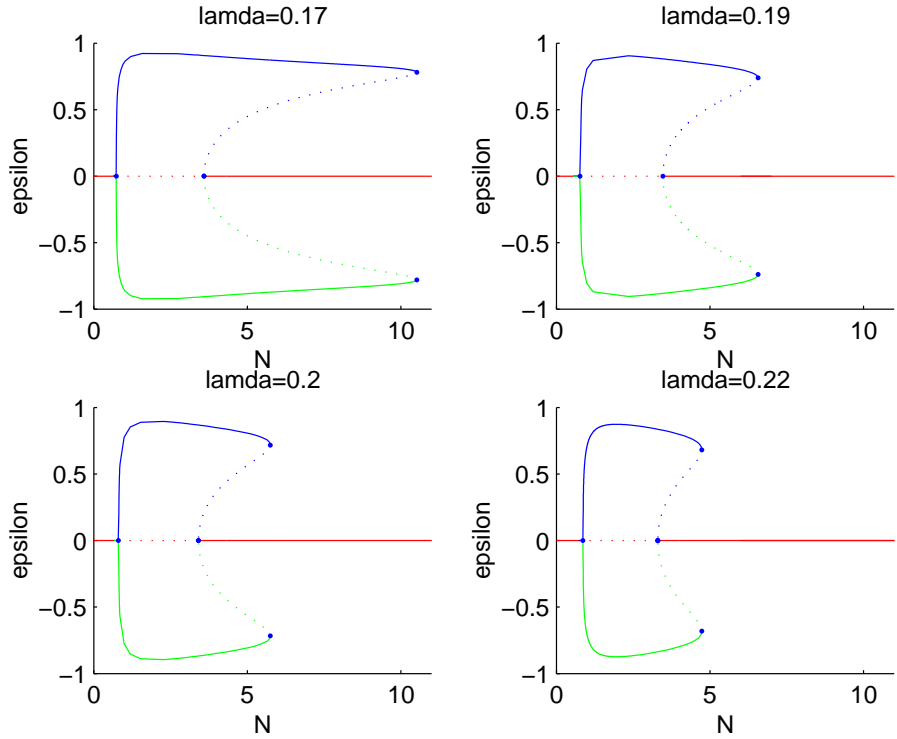}
\caption{A set of bifurcation diagrams for symmetric and asymmetric
solitons, in the plane $\left(N,\protect\epsilon \right) $, as found from
numerical solution of Eqs. (\protect\ref{U}) and (\protect\ref{V}) at
different values of the linear-coupling constant $\protect\lambda $. Stable
and unstable branches of the solutions are shown by solid and dashed curves,
respectively, and bold dots indicate bifurcation points.}
\label{bifurcation1}
\end{figure}
\begin{figure}[tbp]
\includegraphics[width=5.4in]{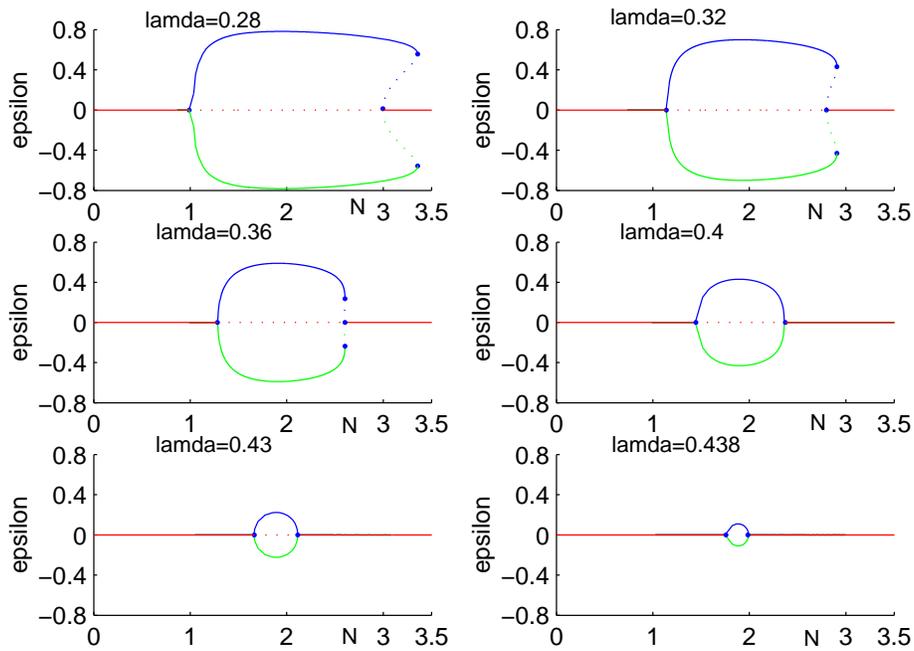}
\caption{Continuation of Fig. \protect\ref{bifurcation1}.}
\label{bifurcation2}
\end{figure}

As expected, Figs. \ref{bifurcation1} and \ref{bifurcation2} demonstrate
\textit{bifurcation loops}. The loop exists at $\lambda \leq \lambda _{\max
}\approx 0.44$ (in fact, the full loop was found for $\lambda \geq 0.10$, as
for smaller $\lambda $ it was difficult to close the loop, due to its very
large size; indeed, for $\lambda \rightarrow 0$, the loop extends to $%
N\rightarrow \infty $, as one then has an obvious stable asymmetric
solution, with $u\neq 0$ and $v=0$, for any $N$). The direct
(symmetry-breaking) bifurcation is observed to be always supercritical,
while the reverse one, which closes the loop, is subcritical (giving rise to
the bistability and concave shape of the loop, on its right side) up to $%
\lambda \approx 0.40$. In a narrow interval of $0.40<\lambda <0.44$, the
reverse bifurcation is supercritical, and the (small) loop has a convex form.

It is relevant to mention that a somewhat similar sequence of bifurcation
loops was reported for (modulationally unstable)\ CW states, rather than
solitons, in a symmetric dual-core model by Snyder \textit{et al}. \cite%
{Snyder} [see Figs. 10(a) and 10(b) in that paper]. However, that sequence
was found in a model with a saturable, rather than cubic-quintic,
nonlinearity in each core.

\begin{figure}[tbp]
\includegraphics[width=4.5in]{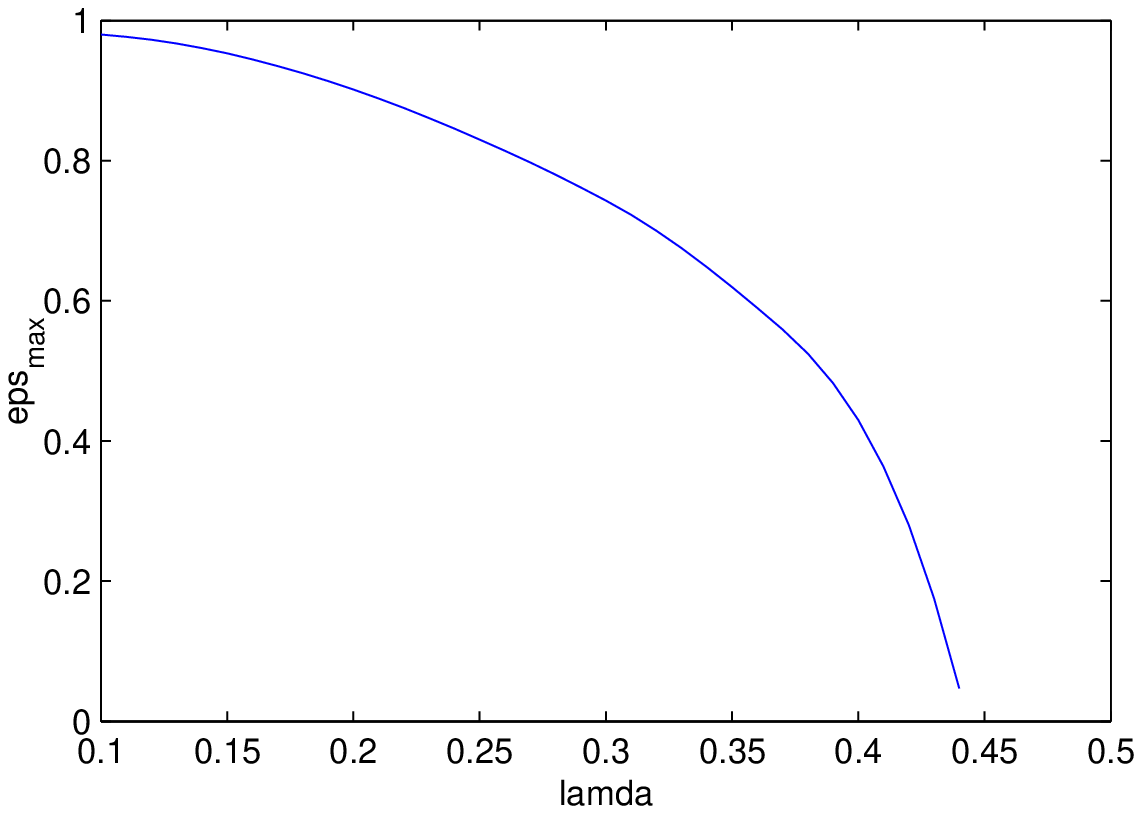}
\caption{The maximum value of the asymmetry parameter (\protect\ref{epsilon}%
) of solitons generated by the bifurcation, versus the linear-coupling
constant $\protect\lambda $. For very small values of $\protect\lambda $, $%
\protect\epsilon _{\max }$ is not shown because of difficulties with
convergence of the numerical scheme.}
\label{epsilon_max}
\end{figure}

\begin{figure}[tbp]
\includegraphics[width=5.4in]{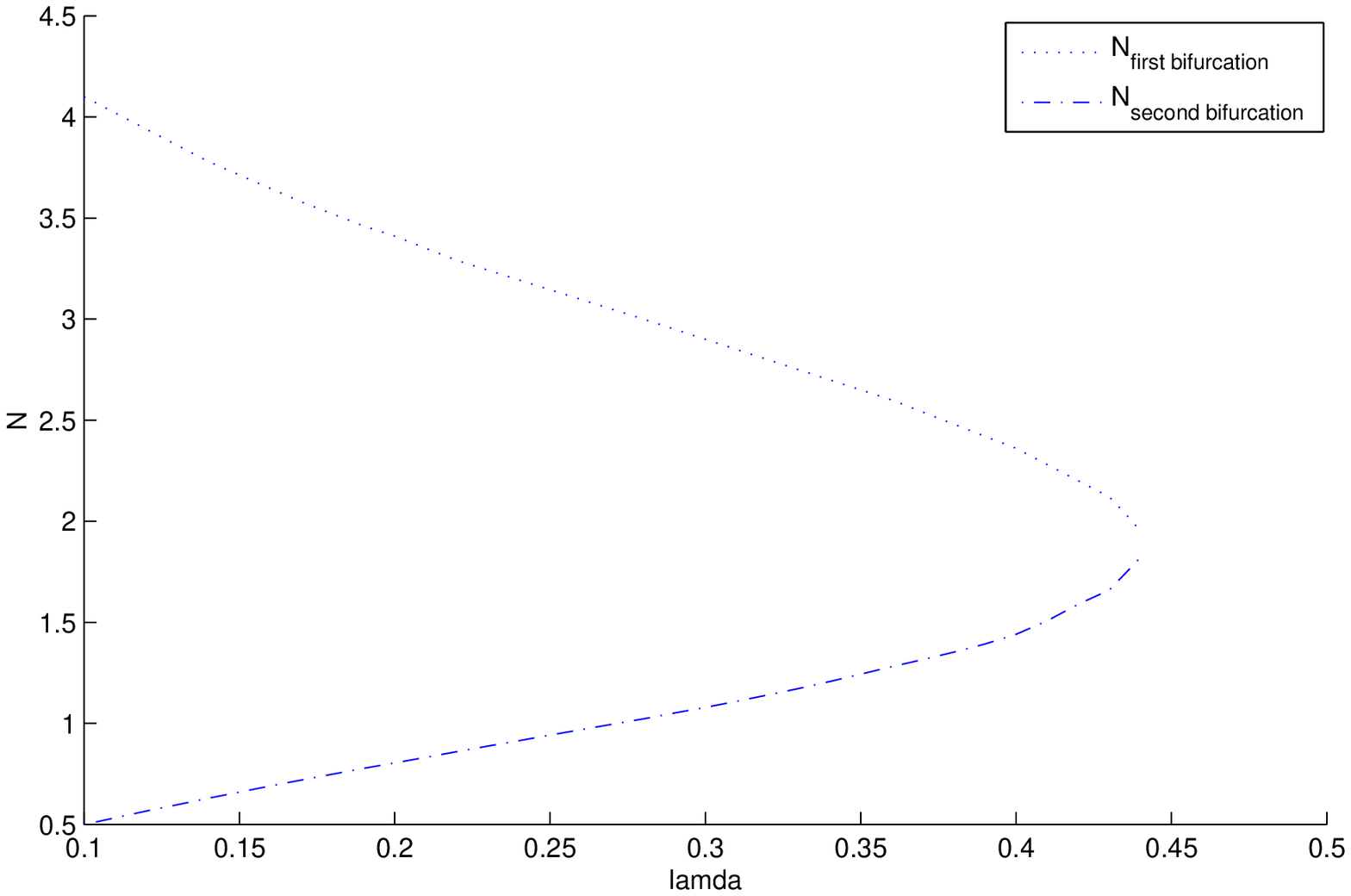}
\caption{Values of the norm (energy) of the symmetric soliton at which the
direct and reverse bifurcations occur. The two curves merge and terminate at
$\protect\lambda =\protect\lambda _{\max }\approx 0.44$.}
\label{Nbif}
\end{figure}

The global picture of the bifurcations is additionally illustrated by Figs. %
\ref{epsilon_max} and \ref{Nbif}. They display, versus $\lambda $, the value
of the asymmetry parameter (\ref{epsilon}) for the most asymmetric soliton
generated by the bifurcation for given $\lambda $, and values of norm $N$ of
the symmetric soliton at points of the direct and reverse bifurcations.

Stability and instability of different branches of soliton solutions can be
anticipated on the basis of general principles of the bifurcation theory
\cite{bif}: the symmetric solution becomes unstable after the direct
supercritical bifurcation, and asymmetric solutions emerge as stable ones at
this point; after the reverse bifurcation, the symmetric solution is stable
again. In the case when the reverse bifurcation is subcritical and,
accordingly, the bifurcation loop is concave on its right side, two branches
of asymmetric solutions (one stable and one unstable) meet and terminate at
each turning point through the saddle-node (alias tangent) bifurcation;
accordingly, the branches of the asymmetric solitons which terminate at the
reverse-bifurcation point are unstable.

These expectations are fully borne out by direct numerical simulations. In
particular, the simulations clearly demonstrate that a small
symmetry-breaking perturbation triggers spontaneous rearrangement of
presumably unstable symmetric solitons into stable asymmetric ones. The
transition (not shown here) is accompanied by relatively small radiative
loss.

Simulations of the evolution of perturbed unstable solitons, corresponding
to the intermediate branches in Figs. \ref{bifurcation1} and \ref%
{bifurcation2} (i.e., in the case of the bistability), demonstrate that the
unstable soliton, which has a choice to transform itself into either a still
more asymmetric soliton, or a symmetric one, both of which are stable,
clearly follows the latter route, even if the initial perturbation acted in
the opposite direction, trying to make the soliton less asymmetric (not
shown here in detail). In this case, the transition entails a larger
radiative loss.

The instabilities of symmetric and asymmetric solitons are qualitatively
different: as said above, the symmetric solutions (\ref{Pushk}) are stable
according to the VK criterion, i.e., their instability may only be
oscillatory [accounted for by complex eigenvalue(s)], which is indeed
observed in the simulations (the validity of the VK criterion was not proved
for the present model, but it was rigorously derived for another system,
including two linearly coupled equations with the cubic nonlinearity of
opposite signs \cite{Valery}). On the other hand, inspection of dependences $%
N(k)$ for the asymmetric solutions demonstrates that, in all cases when
these solitons are unstable, they have $dN/dk<0$, i.e., they are expected to
be VK-unstable solutions. The latter implies that the instability, being
accounted for by a real eigenvalue, must grow without oscillations. The
simulations support this expectation.

Stability of all solution branches which are shown as stable ones in Figs. %
\ref{bifurcation1} and \ref{bifurcation2} has also been verified in direct
simulations. In all these cases, it was found that even relatively large
perturbations do not destroy the solitons.

\section{Collisions between stable solitons}

The underlying system of Eqs. (\ref{u}) and (\ref{v}) is Galilean invariant:
a ``shove" transformation
\begin{equation}
\{u(\tau ),v(\tau )\}\rightarrow \{u(\tau ),v(\tau )\}e^{-i\chi \tau }
\label{shove}
\end{equation}
with an arbitrary real constant $\chi $ generates a \textit{boosted}
solution differing from the original one by $\tau \rightarrow \tau +2\chi z$
, i.e., it moves at the (inverse) velocity $c=-2\chi $ relative to the
original state. The invariance suggests to study collisions between moving
solitons (collisions between symmetric and asymmetric solitons in the model
of a coupler with the cubic nonlinearity were investigated in Ref. \cite%
{Scripta}). Simulations of the collisions make it possible to draw the
following general conclusions.

\begin{figure}[tbp]
\includegraphics[width=5.4in]{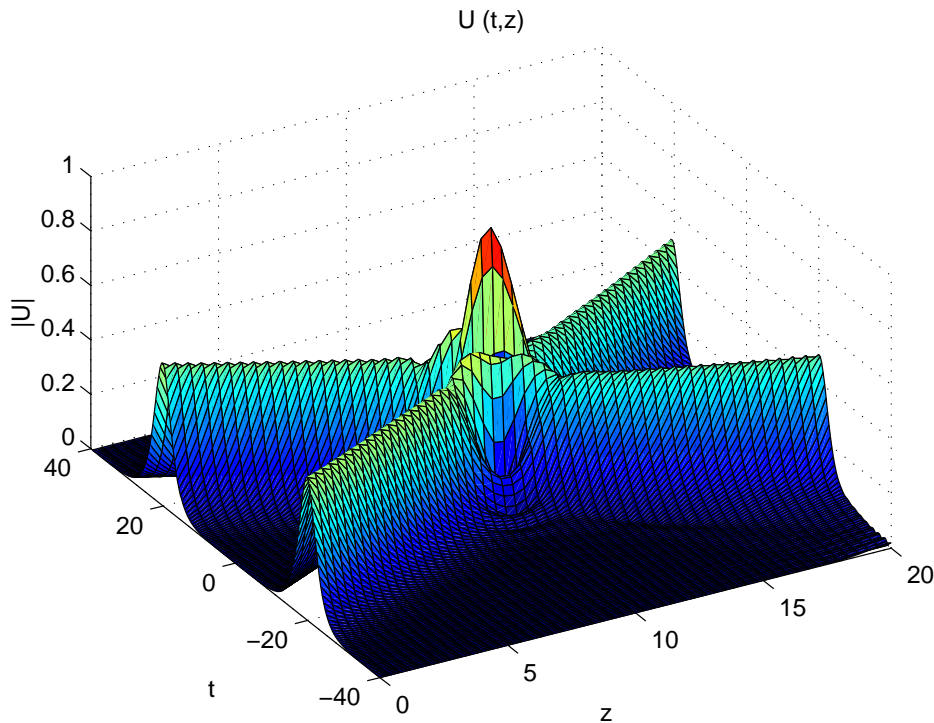}
\caption{A typical example of the elastic collision between two
identical in-phase symmetric solitons generated by means of
expressions (\protect\ref{shove}) with $\protect\chi =\pm 1$, for
$\protect\lambda =0.32$ and $N=1.106 $ (which correspond to
$k=0.56$). Only the $u$-component is shown here, as the picture in
the $v$-component is identical.} \label{elastic-collision}
\end{figure}

First, collisions between stable symmetric solitons (which are
obviously tantamount to collisions between solitons in the
single-component NLS equations with the CQ nonlinearity) appear to
be completely elastic, as shown in Fig. \ref{elastic-collision}.
The elastic collisions give rise to small shifts of the solitons'
centers, in the same fashion as in collisions between solitons in
the integrable cubic NLS equation (each soliton is additionally
shifted in the direction of its motion).

Second, collisions between stable asymmetric solitons are essentially \emph{%
inelastic}. In this case, the solitons emerge from the collision
with smaller velocity $c$, and in an excited state, i.e., as
\textit{breathers}, rather than stationary solitons. For example,
the (inverse) velocities of both solitons drop from
$c_{\mathrm{in}}=\pm 0.4$ before the collision to
$c_{\mathrm{out}}= \pm 0.2$ after the collision, if the moving
solitons were generated by means of
expression (\protect\ref{shove}) with $\protect\chi =\pm 0.2$, for $\protect%
\lambda =0.3$ and $N=1.176$ (which correspond to $k=0.6$).

\begin{figure}[tbp]
\includegraphics[width=5.4in]{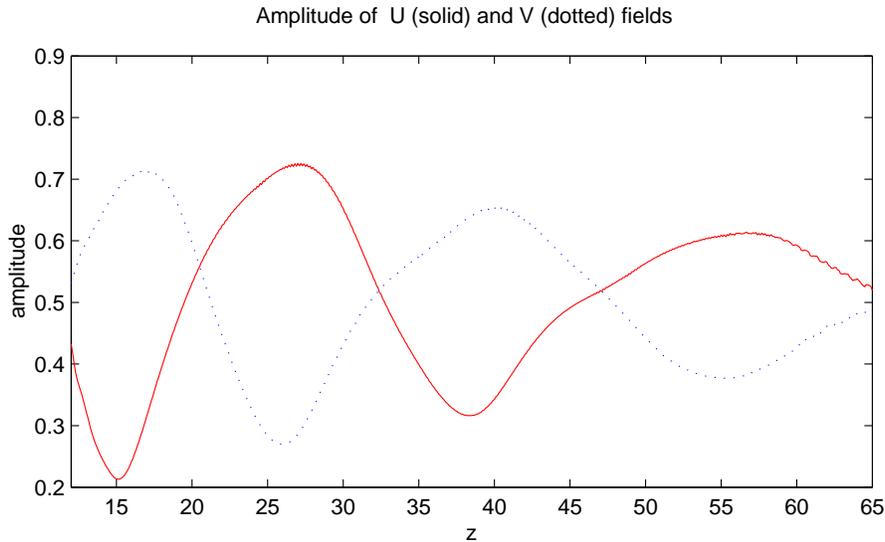}
\caption{The evolution of one of the breathers produced by the
inelastic collision of two solitons in the above-mentioned
example, with $\lambda =0.3$, $N=1.176$, $k=0.6$, and shove
parameters $\chi=\pm 0.2$ (the second breather is a mirror image
of the one shown here). The solid and dashed curves show the
amplitudes of the $u$- and $v$-components of the breather.}
\label{breather}
\end{figure}

The breathers which appear after the inelastic collision are not persistent
ones. As Fig. \ref{breather} shows, the amplitude of the intrinsic
oscillations of the breather gradually decreases, while the oscillation
period increases. This may imply that, in the course of subsequent extended
evolution, the breathers will eventually relax to stationary solitons (due
to losses induced by emission of radiation), but detailed study of this
stage of the evolution is beyond the scope of this work. It is noteworthy
that, while the solitons generally keep their identity after the inelastic
collision (the solitons which originally had a larger $u$- or $v$-component
keep this difference after the collision), the post-collision oscillatory
evolution clearly tends to \emph{symmetrize} both pulses, so they may
eventually settle down to a pare of identical symmetric solitons.

\section{Conclusion}

In this work, we have extended the model of the nonlinear dual-core coupler
by adding defocusing quintic nonlinearity to the usual cubic self-focusing
term. In terms of optics, the model may be interpreted in temporal and
spatial domains alike (i.e., in fiber couplers and in a pair of
tunnel-coupled parallel planar waveguides). As a result, at not too large
values of the linear-coupling parameter $\lambda $, we observe a bifurcation
loop for solitons: the supercritical bifurcation, which destabilizes the
symmetric soliton and creates a pair of asymmetric ones, is followed by a
reverse subcritical bifurcation that restores the stability of the symmetric
soliton. In this case, the loop has the concave shape on its right side, and
includes a large bistability region, which may be of interest for
applications to all-optical switching. At larger values of $\lambda $, the
loop's shape becomes convex, and the bistability disappears. At still larger
$\lambda $, the symmetric solitons always remain stable, and asymmetric ones
never appear.

The stability of all the branches of the soliton states, verified by direct
simulations, exactly complies with what may be predicted, on the basis of
the shape of the bifurcation diagrams, by general principles of the
bifurcation theory. Additionally, in those cases when the asymmetric
solitons are unstable, their instability is correctly predicted by the
Vakhitov-Kolokolov criterion (while the instability of symmetric solitons
has a different nature, as they are unstable against oscillatory
perturbations). The simulations also demonstrate that unstable symmetric
solitons rearrange themselves into stable asymmetric ones, with small
radiation loss. Unstable asymmetric solitons rearranged into their still
more asymmetric stable counterparts.

The Galilean invariance of the model makes it possible to study collisions
between moving solitons. Symmetric solitons collide in a completely elastic
fashion, while collisions between asymmetric solitons are inelastic: they
lose, roughly, half of the velocity, and emerge from the collisions as
breathers, that subsequently tend to symmetrize themselves and, possibly,
settle down to stationary symmetric solitons.

It is plausible that basic results reported in this work are common to
solitons in double-core systems with various forms of saturable
nonlinearity. It may also be quite interesting to consider similar problems
for two-dimensional solitons, that can be realized as spatiotemporal ``light
bullets" \cite{review} in parallel-coupled planar waveguides with the
intrinsic nonlinearity of the cubic-quintic type.

\section*{Appendix: Absence of asymmetric complex solitons generated by a
bifurcation from the symmetric soliton}

Here we aim to consider in more detail a possibility of the existence of
stationary asymmetric\ soliton solutions in the complex form, as per Eqs. (%
\ref{complex}). Substituting the latter in the stationary version of Eqs. (%
\ref{u}) and (\ref{v}), one can easily derive the following pair of
equations,
\begin{equation}
\frac{d}{d\tau }\left(a^{2}\dot{\phi}\right) =\lambda ab\sin \left(\phi
-\psi \right) ,~\frac{d}{d\tau }\left(b^{2}\dot{\psi}\right) =-\lambda
ab\sin \left(\phi -\psi \right) ,  \label{phase}
\end{equation}
and another pair of equations, that we do not need here in an explicit form.
An obvious consequence of Eqs. (\ref{phase}) for localized solutions, with $%
a(|\tau |=\infty )=b(|\tau |=\infty )=0$, is relation (\ref{ab}).

If there is a bifurcation from the real symmetric soliton (\ref{Pushk}) that
gives rise to complex asymmetric solitons with a nontrivial phase structure,
then, in an infinitesimal proximity to the bifurcation point, where both the
phase functions $\phi (\tau )$ and $\psi (\tau )$ and deviations of $a(\tau
) $ and $b(\tau )$ from expression (\ref{Pushk}) are infinitely small, Eq. (%
\ref{ab}) yields, in the lowest-order approximation, $\dot{\phi}+\dot{\psi}
=0 $, or, in other words, $\psi(\tau)=-\phi (\tau )$ (it is trivial to
remove a phase constant). Further, in the same approximation, an equation
for an infinitesimal function $\phi (\tau )$ following from Eqs. (\ref{phase}%
) is
\begin{equation}
-\frac{d^{2}\phi }{dT^{2}}+2\lambda a^{4}(T)~\phi =0,  \label{quantumMech}
\end{equation}
where the expression (\ref{Pushk}) is to be substituted for $a$, and
\begin{equation}
T\equiv \int_{0}^{\tau }\frac{d\tau ^{\prime }}{a^{2}(\tau ^{\prime })}=
\frac{1}{2\left(k-\lambda \right) }\left[ \tau +\sqrt{\frac{1}{%
4\left(k-\lambda \right) }-\frac{1}{3}}\sinh \left(2\sqrt{k-\lambda }\tau
\right) \right]  \label{T}
\end{equation}
($T$ is a monotonous function of $\tau $, varying together with it from $%
-\infty $ to $+\infty $). Equation (\ref{quantumMech}) with $\lambda >0$ is
a linear Schr\"{o}dinger equation with a coordinate $T$ and potential \emph{%
\ barrier} (rather than a potential well), which, obviously, gives rise to
no localized eigenfunctions, hence it cannot generate any bifurcation that
would be signalled by the appearance of such a solution (if one sets $%
\lambda <0$, the conclusion will be the same, as in this case one will start
with the antisymmetric soliton).

\end{document}